\documentclass[aps,showpacs,twocolumn,superscriptaddress]{revtex4}
\usepackage[dvips]{graphicx}
\usepackage{amsmath}
\usepackage{color}

\begin{document}

\title{Heat conduction and energy diffusion in momentum-conserving 1D full lattice ding-a-ling model}

\author{Zhibin Gao}
\affiliation{Center for Phononics and
Thermal Energy Science and School of Physics Science and
Engineering, Tongji University, 200092 Shanghai, People's Republic
of China}

\author{Nianbei Li}
\email{nbli@tongji.edu.cn} \affiliation{Center for Phononics and
Thermal Energy Science and School of Physics Science and
Engineering, Tongji University, 200092 Shanghai, People's Republic
of China}

\author{Baowen Li}
\email{phononics@tongji.edu.cn} \affiliation{Department of Mechanical Engineering, University of Colorado Boulder, CO 80309, USA}


\begin{abstract}
The ding-a-ling model is a kind of half lattice and half hard-point-gas (HPG) model. The original ding-a-ling model proposed by Casati {\it et.al} does not conserve total momentum and has been found to exhibit normal heat conduction behavior. Recently, a modified ding-a-ling model which conserves total momentum has been studied and normal heat conduction has also been claimed. In this work, we propose a full lattice ding-a-ling model without hard point collisions  where total momentum is also conserved. We investigate the heat conduction and energy diffusion of this full lattice ding-a-ling model with three different nonlinear inter-particle potential forms. For symmetrical potential lattices, the thermal conductivities diverges with lattice length and their energy diffusions are superdiffusive signaturing anomalous heat conduction. For asymmetrical potential lattices, although the thermal conductivity seems to converge as the length increases, the energy diffusion is definitely deviating from normal diffusion behavior indicating anomalous heat conduction as well. No normal heat conduction behavior can be found for the full lattice ding-a-ling model.
\end{abstract}

\maketitle
\section{Introduction}
It has long been believed that there should be profound connection between heat conduction and energy diffusion since both of them describe the same
dynamic process of energy transport~\cite{Helfand1960pr}. For normal and ballistic heat conduction, it is well established that their energy diffusions should also be normal and ballistic,
respectively. What is nontrivial and really interesting is the connection between anomalous heat conduction and anomalous energy diffusion for low dimensional systems.
For anomalous heat conduction, the thermal conductivity diverges with the system length as $\kappa\propto N^{\alpha}$ with $0<\alpha<1$~\cite{Lepri2003pr,Dhar2008ap,Wang2008epjb,Liu2013epjb,Lepri1997prl,Lepri1998pre,Hatano1999pre,Narayan2002prl,Grassberger2002prl,Maruyama2002p,
Segal2003jcp,Wang2004prl,Li2005chaos,Zhang2005jcp,Pereira2006prl,Chang2008prl,Henry2008prl,Henry2009prb,Yang2010nt,Wang2011epl,Beijeren2012prl,Xiong2012pre,Zhong2012pre,
Wang2012pre,Liu2012prb,Wang2013pre,Mendl2013prl,Savin2014pre,Das2014pre,Mendl2014pre,Spohn2014jsp,Xiong2014pre,Xu2014nc}. For anomalous energy diffusion,
the spreading of the Mean Square Displacement (MSD) of energy follows the behavior as $\left<\Delta x^2(t)\right>_E\propto t^{\beta}$ with $1<\beta<2$~\cite{Zhao2006prl,Cipriani2005prl,Delfini2007epjst}.
In the early pioneering studies, it has been found that the connection formula between heat conduction and energy diffusion is $\alpha=\beta-1$~\cite{Zhao2006prl,Cipriani2005prl,Delfini2007epjst,Denisov2003prl,Li2010prl}. Only until
recently, a rigorous connection theory between heat conduction and energy diffusion has been established within the framework of linear response theory~\cite{Liu2014prl}. As a
byproduct, the above mentioned connection formula comes out quite naturally.

According to the connection theory~\cite{Liu2014prl}, the study of the energy diffusion is equivalent to the study of heat conduction in the sense of calculating thermal conductivity.
As has been pointed out by Ref. \cite{Li2014arxiv}, the temporo-spatial distribution of energy fluctuation correlation function gives more information than the heat current-current
correlation function used in Green-Kubo formula. This justifies the advantage of using energy diffusion method to study the problem of heat conduction.
In previous works, it has been found that the energy fluctuation correlation function is Gaussian for systems with normal heat conduction~\cite{Zhao2006prl}. More interestingly, the energy
fluctuation correlation function for systems with anomalous heat conduction is a Levy-walk like distribution~\cite{Zhao2006prl,Cipriani2005prl,Delfini2007epjst,Zaburdaev2011prl,Das2014pre,Mendl2014pre,Spohn2014jsp,Li2014arxiv}. For a comprehensive understanding of the Levy walks, please refer to the following excellent references~\cite{Klafter1993p,Zaburdaev2014arxiv}. In general, the 1D lattice system exhibits
normal heat conduction behavior if the total momentum is not conserved and anomalous heat conduction behavior if the total momentum is conserved~\cite{Lepri2003pr,Dhar2008ap,Wang2008epjb,Liu2013epjb}. There is
however one exception for the 1D coupled rotator model which conserves total momentum but shows normal heat conduction behavior~\cite{Giardina2000prl,Gendelman2000prl}. Most recently, it has been confirmed that the 1D coupled rotator model possesses normal momentum diffusion as well as normal energy diffusion~\cite{Li2014arxiv,Spohn2014arxiv,Dhar2014arxiv}.

The first 1D ding-a-ling model proposed to study the heat conduction problem was introduced by Casati {\it et.al} in 1984~\cite{Casati1984prl}. This system consists of on-site harmonic
oscillators and free moving particles positioned alternately. It is a kind of half lattice and half HPG model where total momentum is not conserved.
Therefore it is not a surprise that normal heat conduction has been found for this model~\cite{Casati1984prl}. Recently, a modified ding-a-ling model was proposed where the on-site harmonic oscillators are replaced by inter-connected harmonic oscillators~\cite{Dadswell2010pre}.
With this setup, the total momentum is conserved. It is then quite surprising that normal heat conduction has also been detected for this modified half lattice and half HPG ding-a-ling model~\cite{Dadswell2010pre}.

In this paper, we will propose a full lattice ding-a-ling model which conserves the total momentum. Three different inter-particle potentials will be applied to this model. The heat conduction and energy diffusion behaviors will be studied for each potential form with two numerical methods. Firstly, the length dependent thermal conductivities $\kappa$ will be calculated using the non-equilibrium molecular dynamics simulations. Secondly, the energy fluctuation correlation function will be calculated at different correlation times and the MSD of energy spreading will be presented to reveal the heat conduction behavior for this momentum conserving full lattice ding-a-ling model. This paper will be organized as the followings: in Sec. II we will give the introduction of the full lattice ding-a-ling model with three different inter-particle potentials. The heat conduction and energy diffusion behaviors will be presented in Sec. III. In Sec. IV, we will give our summary for the study of heat conduction as well as energy diffusion for our new proposed full lattice ding-a-ling model.

\begin{figure}[t]
\includegraphics[width=\columnwidth]{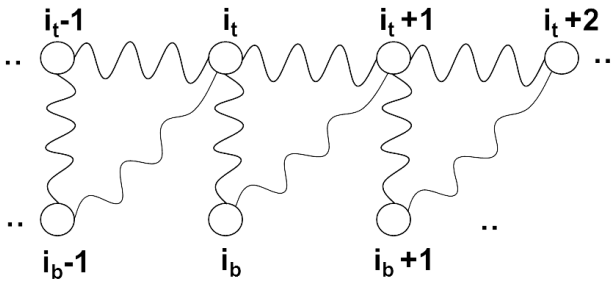} 
\vspace{-0.5cm} \caption{\label{fig:sketch} 
The schematic setup for the full lattice ding-a-ling model conserving total momentum. The top particles are connected by harmonic springs. The bottom particles are connected with its
two neighbors of top particles with symmetrical and asymmetrical nonlinear inter-particle potentials.}
\end{figure}

\section{1D momentum conserving full lattice ding-a-ling models}
The full lattice ding-a-ling model is depicted in Fig. \ref{fig:sketch}. The original half free moving particles shown as the bottom particles in Fig. \ref{fig:sketch} are now connected with its two top neighbors via the nonlinear inter-particle potentials. The Hamiltonian for this full lattice ding-a-ling model is the
following:
\begin{eqnarray}\label{ham}
H=\sum_{i}H_i=\sum_{i_t,i_b}\left[\frac{p^2_{i_t}}{2}+\frac{p^2_{i_b}}{2}+\frac{1}{2}(q_{i_t+1}-q_{i_t})^2\right.\nonumber\\
 \left.+V(q_{i_b},q_{i_t})+V(q_{i_b},q_{i_t+1})\right]
\end{eqnarray}
where the index $i_t$ and $i_b$ denote the particle index in the top and bottom respectively. The $p_{i_t}$ and $p_{i_b}$ are the momentum for top and bottom
particles. The $q_{i_t}$ and $q_{i_b}$ are the displacements from equilibrium for top and bottom particles.

\begin{figure}[t]
\includegraphics[width=0.65\columnwidth]{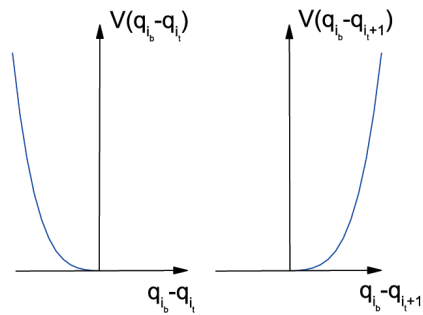}
\vspace{-0.25cm} \caption{\label{fig:half-fpu}
(Color online) The schematic picture of the asymmetrical half FPU-$\beta$ form potential $V(q_{i_b}-q_{i_t})$ and $V(q_{i_b}-q_{i_t+1})$.}
\end{figure}

The inter-particle potential $V(q_{i_t},q_{i_b})$ describes the interaction between top and bottom particles which could be any nonlinear potential form. Here we consider three typical nonlinear potentials for $V$: (i) symmetric FPU-$\beta$
potential with
\begin{equation}
V(q_{i_b},q_{i_t})=\frac{(q_{i_b}-q_{i_t})^2}{2}+\frac{(q_{i_b}-q_{i_t})^4}{4};
\end{equation}
(ii) symmetric coupled rotator potential with
\begin{equation}
V(q_{i_b},q_{i_t})=1-\cos(q_{i_b}-q_{i_t});
\end{equation}
(iii) asymmetric half FPU-$\beta$ potential depicted in Fig. \ref{fig:half-fpu} with
\begin{eqnarray}
&&V(q_{i_b},q_{i_t})\nonumber\\
&&=\left\{
\begin{aligned}
\frac{(q_{i_b}-q_{i_t})^2}{2}+\frac{(q_{i_b}-q_{i_t})^4}{4}, \,\,q_{i_b}-q_{i_t}<0 \\
 0, \,\,q_{i_b}-q_{i_t}\ge 0
\end{aligned}
\right.\\
&&V(q_{i_b},q_{i_t+1})\nonumber\\
&&=\left\{
\begin{aligned}
0,q_{i_b}-q_{i_t+1}<0 \\
\frac{(q_{i_b}-q_{i_t+1})^2}{2}+\frac{(q_{i_b}-q_{i_t+1})^4}{4}, q_{i_b}-q_{i_t+1}\ge 0\nonumber
\end{aligned}
\right.
\end{eqnarray}

The first two models are symmetrical since $V(q_{i_b},q_{i_t})$ and $V(q_{i_b},q_{i_t+1})$ posses the same form while the third
model is asymmetric as $V(q_{i_b},q_{i_t})$ and $V(q_{i_b},q_{i_t+1})$ have different potential forms. Compared with the previously studied half lattice and half HPG ding-a-ling model, this full lattice ding-a-ling model enables us to simulate
the system dynamics with more accurate and consistent numerical algorithm such as fourth order symplectic method~\cite{Laskar2001} in studying the energy diffusion behaviors. In the following calculation, the dimensionless units have been applied.

\begin{figure}[t]
\includegraphics[width=\columnwidth]{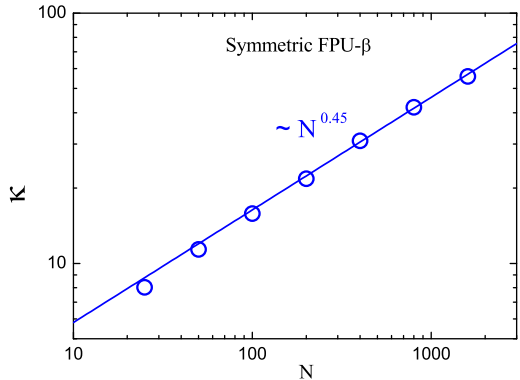}
\vspace{-0.5cm} \caption{\label{fig:sym-fpub_kappa}
(color online). Thermal conductivity $\kappa$ as the function of lattice size $N$ for symmetric FPU-$\beta$ potential. The circles are the numerical data and the straight line of $\kappa\propto N^{0.45}$ is guided for your eyes. The average temperature is chosen as $T=12.15$ corresponding to energy density $E=10$ used in equilibrium method below. Temperature bias is set as $\Delta=0.1$ and $N$ is the number of unit cells. }
\end{figure}

\section{Heat Conduction and Energy Diffusion}
In the study of heat conduction, we first apply the direct non-equilibrium molecular dynamics method. For simplicity and convenience, the fixed boundary conditions with $q_{i_t=0}=q_{i_t=N+1}=0$ will be used. The first and last atoms on the top chain will be contacted with two Langevin heat baths with temperature $T_{L/R}$. In particular, the equations of motions for these two atoms are
\begin{equation}
\ddot{q}_{i_t}=F(q_{i_t})-\lambda \dot{q}_{i_t}+\xi_{i_t}(t)
\end{equation}
where $F(q_{i_t})$ is the force which can be derived from Eq. (\ref{ham}) and corresponding boundary conditions, and the Gaussian white noise $\xi_{i_t=1/N}(t)$ satisfy
\begin{eqnarray}
\left<\xi_{i_t=1/N}(t)\right>&=&0\nonumber\\
\left<\xi_{i_t=1/N}(t)\xi_{i_t=1/N}(0)\right>&=&2\lambda k_{B}T_{L/R}\delta(t)
\end{eqnarray}
where $\lambda$ is the damping parameter which will be always set as unity and $\left<\cdot\right>$ denotes the ensemble average which is equivalent
to the time average for the chaotic systems we considered here. The second order Verlet velocity method will be used to integrate the equations of motions.

The unit cell contains one top particle $i_t$ and one bottom particle $i_b$ indexed by $i=i_t=i_b$. From continuity equation of energy for this unit cell, the heat flux can be derived as $j_i=-\dot{q}_{i_t}(q_{i_t}-q_{i_t-1})-\dot{q}_{i_t}\partial{V(q_{i_b-1},q_{i_t})}/\partial{q_{i_t}}$. The thermal conductivity can be calculated as
\begin{equation}
\kappa=-\frac{J}{\nabla{T}}
\end{equation}
where $J=\left<j_i\right>$ is the average heat flux in the stationary state independent on the index $i$ and $\nabla{T}$ is the temperature gradient. The temperature of the heat baths are set as $T_{L/R}=T(1\pm\Delta)$ where $T$ is the average temperature and $\Delta$ is the bias.

In the study of energy diffusion, the equilibrium numerical method will be applied with the periodic boundary conditions of $q_{i_t}=q_{N+{i_t}}$ and $q_{i_b}=q_{N+{i_b}}$. The key information needed to be calculated is the energy fluctuation correlation function $C_E(i,t)$ defined as \cite{Zhao2006prl}
\begin{equation}\label{CeCp}
\begin{split}
C_E(i,t)&=\frac{\langle \Delta H_i(t)\Delta H_0(0) \rangle}{\langle \Delta H_0(0)\Delta H_0(0)\rangle}
\end{split}
\end{equation}
where $\Delta H_i(t)\equiv H_i(t)-\langle H_i \rangle$. The local energy $H_i$ includes the energy of
top particle $i_t$ and bottom particle $i_b$ with $i=i_t=i_b$. The unit cell index $i$ runs from $-(N-1)/2$ to $(N-1)/2$ so there are $N$ top and bottom particles each. With this indexing for odd $N$, the central particles have the index $i_t=i_b=0$ which turns out to be convenient for numerical simulations. The system will have $2N$ particles in total. Note that $C_{E}(i,t=0)=\delta_{i,0}$ is the initial distribution for energy fluctuation correlation function. If one consider a small initial excess energy perturbation with a special delta form, the excess energy distribution $\rho_E(i,t)$ equals to the energy fluctuation correlation function as $\rho_E(i,t)=C_E(i,t)$, which describes
the actual time evolution of the initial excess energy along the lattice~\cite{Liu2014prl}.

\begin{figure}[t]
\includegraphics[width=\columnwidth]{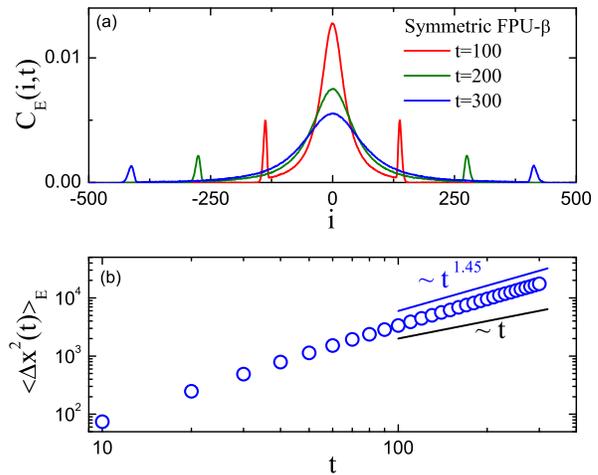}
\vspace{-0.5cm} \caption{\label{fig:sym-fpub}
(color online). (a) Energy fluctuation correlation function $C_E(i,t)$ for model with symmetric FPU-$\beta$ potential. The red, olive and
blue solid lines represent the functions at $t=100$, $200$ and $300$, respectively. (b) The MSD of the excess energy $\left<\Delta x^2(t)\right>_E$
as the function of correlation time $t$. The two solid reference lines of $t^{1.45}$ and $t$ are guided for the eyes. It can be seen
that the numerical data almost follows the $t^{1.45}$ behavior which is the similar anomalous energy diffusion behavior previously found for 1D FPU-$\beta$ lattice. The energy density $E=10$ and the lattice length $N=1001$.}
\end{figure}

The MSD of the excess energy distribution $\left<\Delta x^2(t)\right>_E$ can be defined as~\cite{Zhao2006prl,Liu2014prl}:
\begin{equation}
\left<\Delta x^2(t)\right>_E\equiv\sum_{i}i^2\rho_E(i,t)=\sum_{i}i^2 C_E(i,t)
\end{equation}
According to the connection theory, the second derivative of the MSD of the excess energy distribution $\left<\Delta x^2(t)\right>_E$ is connected with the autocorrelation function
of total heat flux $C_{JJ}(t)$~\cite{Liu2014prl}:
\begin{equation}
\frac{d^2\left<\Delta x^2(t)\right>_E}{dt^2}=\frac{2C_{JJ}(t)}{k_B T^2 c}
\end{equation}
where $c$ is the volumetric specific heat capacity and $k_B$ is the Boltzmann constant. The autocorrelation function of total
heat flux $C_{JJ}(t)$ is the central quantity which enters the Green-Kubo formula for the calculation of thermal conductivity~\cite{Lepri2003pr,Dhar2008ap}:
\begin{equation}
\kappa=\frac{1}{k_B T^2}\int^{\infty}_{0}C_{JJ}(t)dt
\end{equation}

The connection theory between heat conduction and energy diffusion tells us the following\cite{Liu2014prl}: if the energy fluctuation correlation function $C_E(i,t)$ is Gaussian in the asymptotic time limit, the MSD of the energy distribution $\left<\Delta x^2(t)\right>_E$ will be linearly proportional to time $t$ as $\left<\Delta x^2(t)\right>_E\propto t$. This linear time dependence of $\left<\Delta x^2(t)\right>_E$ eventually gives rise to a finite
thermal conductivity $\kappa$ indicating a normal heat conduction behavior. This has been already observed for $\phi^4$ lattice~\cite{Zhao2006prl} and coupled rotator lattice~\cite{Li2014arxiv} where both of the systems exhibit normal heat conduction~\cite{Hu2000pre,Aoki2000pla}. If the energy fluctuation correlation function $C_E(i,t)$ is Levy-walk like in the asymptotic time limit,
the MSD will grow faster than linear time dependency as $\left<\Delta x^2(t)\right>_E\propto t^{\beta}$ with $\beta>1$. This is the case for 1D momentum conserving FPU-$\beta$
lattice, amended rotator lattice and Lennard-Jones lattice~\cite{Zhao2006prl,Li2014arxiv}. From the connection theory\cite{Liu2014prl}, the thermal conductivity will diverge as $\kappa\propto N^{\alpha}$ with $\alpha=\beta-1$ exhibiting anomalous heat conduction behavior.

\begin{figure}[t]
\includegraphics[width=\columnwidth]{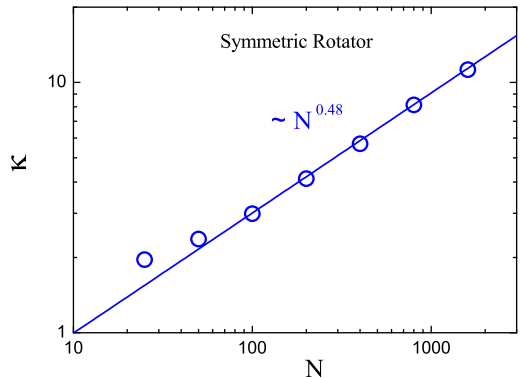}
\vspace{-0.5cm} \caption{\label{fig:sym-rotator_kappa}
(color online). Thermal conductivity $\kappa$ as the function of lattice size $N$ for symmetric rotator potential. The circles are the numerical data and the straight line of $\kappa\propto N^{0.48}$ is guided for your eyes. The average temperature is chosen as $T=0.91$ corresponding to the energy density $E=1$ used in equilibrium method below. Temperature bias is set as $\Delta=0.2$ and $N$ is the number of unit cells.}
\end{figure}

In order to study the heat transport behavior for the full lattice ding-a-ling models with three different nonlinear inter-particle potentials, we will use the two numerical methods to investigate the heat conduction as well as the energy diffusion behaviors. For the non-equilibrium molecular dynamics simulations, the length dependent thermal conductivities $\kappa$ will be calculated directly. For the equilibrium simulations, we will calculate the temporo-spatial distributions of energy fluctuation correlation functions $C_E(i,t)$ at different correlation times. The resulted MSD of the excess energy $\left<\Delta x^2(t)\right>_E$ as the function of correlation time $t$ will be used to analyze the actual behavior of energy diffusion for each model.

\begin{figure}[t]
\includegraphics[width=\columnwidth]{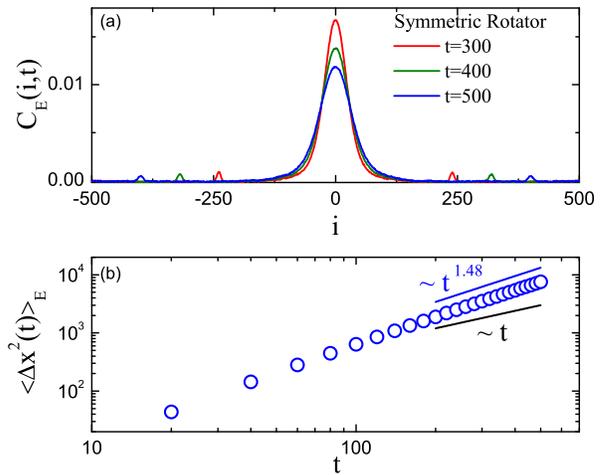}
\vspace{-0.5cm} \caption{\label{fig:sym-rotator}
(color online). (a) Energy fluctuation correlation function $C_E(i,t)$ for model with symmetric rotator potential. The red, olive and
blue solid lines represent the functions at $t=300$, $400$ and $500$, respectively. (b) The MSD of the excess energy $\left<\Delta x^2(t)\right>_E$
as the function of correlation time $t$. The two solid reference lines of $t^{1.48}$ and $t$ are guided for the eyes. It can be seen
that the numerical data also follows the $t^{1.48}$ behavior which is the similar anomalous energy diffusion behavior previously found for 1D FPU-$\beta$ lattice.
The energy density $E=1$ and the lattice length $N=1001$.}
\end{figure}

\subsection{Symmetric FPU-$\beta$ potential}
We first consider the full lattice ding-a-ling model with symmetric FPU-$\beta$ potential. The length dependent thermal conductivities $\kappa$ are plotted in Fig. \ref{fig:sym-fpub_kappa} with average temperature $T=12$. The $\kappa$ shows a good power-law length dependence as $\kappa\propto N^{0.45}$. The heat conduction should be anomalous similar to the FPU-$\beta$ lattice. Simulations with other average temperatures show almost the same results.

In Fig. \ref{fig:sym-fpub}(a), the energy fluctuation correlation functions $C_E(i,t)$ with symmetric FPU-$\beta$ potential have been plotted at different
correlation times $t=100$, $200$ and $300$. Similar to the most studied 1D FPU-$\beta$ lattice~\cite{Zhao2006prl}, the correlation function $C_E(i,t)$ here is also like
the Levy-walk distribution which is characterized by one central peak and two side peaks. The side peaks moves towards outside
with a constant sound velocity. To determine the diffusion behavior, the time dependence of MSD of the excess energy $\left<\Delta x^2(t)\right>_E$
needed to be calculated. The MSD of the excess energy $\left<\Delta x^2(t)\right>_E$ as the function of correlation time $t$ has been plotted in Fig. \ref{fig:sym-fpub}(b).
As can be seen from the figure, the MSD of the excess energy exhibits superdiffusion behavior as $\left<\Delta x^2(t)\right>_E\propto t^{1.45}$. Our
results are consistent with the previously found diffusion exponent $\beta = 1.40$ for the 1D FPU-$\beta$ lattice~\cite{Zhao2006prl}. According to the connection theory
between energy diffusion and heat conduction\cite{Liu2014prl}, superdiffusion of energy of $\beta=1.45$ will imply an anomalous heat conduction with $\kappa\propto N^{\alpha=0.45}$
for the considered full lattice ding-a-ling model with symmetric FPU-$\beta$ potential. This momentum-conserving ding-a-ling model exhibits almost the same energy diffusion behavior
as the 1D momentum-conserving FPU-$\beta$ lattice. We have also done the simulations for this model at other energy densities and similar results are found.

\begin{figure}[t]
\includegraphics[width=\columnwidth]{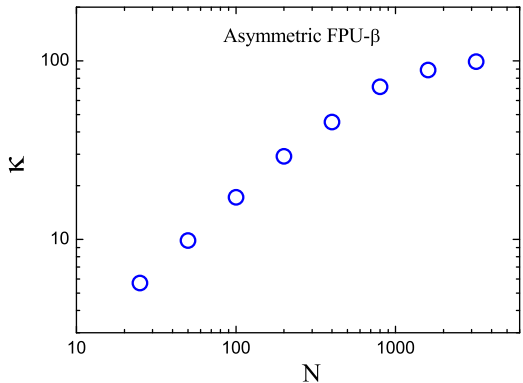}
\vspace{-0.5cm} \caption{\label{fig:asym-fpub_kappa}
(color online). Thermal conductivity $\kappa$ as the function of lattice size $N$ for asymmetric FPU-$\beta$ potential. The circles are the numerical data and the thermal conductivity $\kappa$ seems to converge in the larger $N$ region. The average temperature is chosen as $T=1.0$ corresponding to the energy density $E=1$ used in equilibrium method below. Temperature bias is set as $\Delta=0.2$ and $N$ is the number of unit cells.}
\end{figure}

\subsection{Symmetric rotator potential}
We then consider the full lattice ding-a-ling model with symmetric rotator potential. The length dependent thermal conductivities $\kappa$ are plotted in Fig. \ref{fig:sym-rotator_kappa} with average temperature $T=0.9$. The thermal conductivity $\kappa$ still shows a good power-law length dependence as $\kappa\propto N^{0.48}$. The heat conduction is anomalous as previous case. Simulations with other average temperatures show almost the same results.

In Fig. \ref{fig:sym-rotator}(a), we plot the energy fluctuation correlation functions $C_E(i,t)$ at different correlation times $t=300$, $400$ and $500$ for
the ding-a-ling model with symmetric rotator potential. The clear characteristics of Levy-walk distribution with one central peak and two side peaks appear.
Compared with the ding-a-ling lattice with FPU-$\beta$ potential, the side peaks here are relatively small. But the time dependence of the MSD of the excess energy still
follows the same time dependent behavior as $\left<\Delta x^2(t)\right>_E\propto t^{1.48}$ as can be seen from Fig. \ref{fig:sym-rotator}(b). According to the
connection theory, the heat conduction for the ding-a-ling lattice with rotator potential is anomalous with $\kappa\propto N^{0.48}$ which is the case in Fig. \ref{fig:sym-rotator_kappa}.

It is interesting that although the 1D rotator lattice exhibits normal energy diffusion and normal heat conduction behavior~\cite{Li2014arxiv,Spohn2014arxiv,Dhar2014arxiv}, the ding-a-ling model with rotator potential between top and bottom atoms
displays the similar super energy diffusion and anomalous heat conduction behavior as the 1D FPU-$\beta$ lattice~\cite{Zhao2006prl}.

\begin{figure}[t]
\includegraphics[width=\columnwidth]{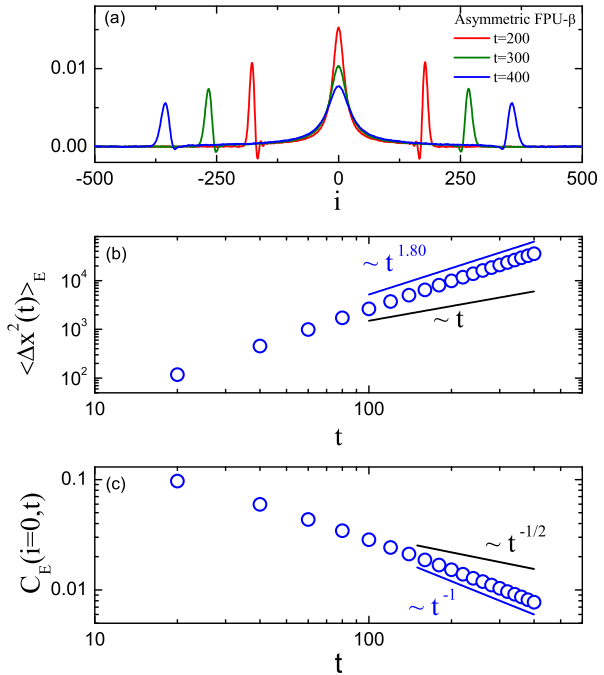}
\vspace{-0.5cm} \caption{\label{fig:asym-fpub}
(color online). (a) Energy fluctuation correlation function $C_E(i,t)$ for model with asymmetric FPU-$\beta$ potential. The red, olive and
blue solid lines represent the functions at $t=200$, $300$ and $400$, respectively. (b) The MSD of the excess energy $\left<\Delta x^2(t)\right>_E$
as the function of correlation time $t$. The two solid reference lines of $t^{1.80}$ and $t$ are guided for the eyes. It can be seen
that the numerical data follows the $t^{1.80}$ behavior which is faster than the anomalous energy diffusion behavior previously found for 1D FPU-$\beta$ lattice. (c) The decay of the central peak $C_E(i=0,t)$. It is obvious that the decay of the central peak as $C_E(i=0,t)\propto t^{-1}$ is faster than the decay for normal diffusion case which should be proportional to $t^{-1/2}$. The energy density $E=1$ and the lattice length $N=1001$.}
\end{figure}

\subsection{Asymmetric FPU-$\beta$ potential}
To further mimic the half lattice and half HPG ding-a-ling model where the free particle can only collide with the particle in its moving direction,
we introduce the ding-a-ling lattice with asymmetric FPU-$\beta$ potential as can be seen in Fig. \ref{fig:half-fpu}. The thermal conductivities $\kappa$ are plotted as the function of length $N$ in Fig. \ref{fig:asym-fpub_kappa}. The $\kappa$ first increases with length $N$ and then seems to saturate to a constant value after $N>1000$. It looks like that $\alpha=0$ and the heat conduction is normal in this asymmetric case.

However, if we look at the energy fluctuation correlation functions $C_E(i,t)$ at different correlation
times $t=200$, $300$ and $400$ plotted in Fig. \ref{fig:asym-fpub}(a), the signatures of Levy-walk distribution with one central peak and two
side peaks have also been found. The only differences are the negative tips next to the side peaks which has already been found for other 1D asymmetric lattices.
The time dependence of the MSD of the excess energy has been plotted in Fig. \ref{fig:asym-fpub}(b) and a superdiffusion with $\left<\Delta x^2(t)\right>_E\propto t^{1.80}$
has been obtained. The relation between $\alpha$ and $\beta$ here deviates from the connection theory as noticed in recent work \cite{Wang2015pre}. It might be that the asymptotic length is very large for asymmetric lattices.

We then further plot the decay of the central peak $C_E(i=0,t)$
in Fig. \ref{fig:asym-fpub}(c). It is observed that the decay rate is proportional to $t^{-1}$ which is also much faster than the decay of the central peak for a normal diffusion which should be proportional to $t^{-1/2}$. Therefore, it will be insufficient to judge the heat conduction behavior only via the non-equilibrium molecular dynamics simulation for asymmetric potential lattices. One should rely on the energy diffusion behavior as well.

\section{Summary}
In summary, we have systematically studied the heat conduction and energy diffusion in momentum-conserving 1D full lattice
ding-a-ling model with symmetric and asymmetric potentials. For symmetric lattices, the thermal conductivities diverge as the lattice length with a power-law dependence displaying obvious anomalous behaviors. The super energy diffusions of $\left<\Delta x^2(t)\right>_E\propto t^{\beta}$
with $\beta>1$ have also been found. For asymmetric lattices, the thermal conductivity seems to converge as the lattice length increases. However, the energy diffusion behavior is definitely superdiffusive indicating anomalous heat conduction behavior also. As a result, the heat conduction for the full lattice ding-a-ling models should be anomalous as that in
1D FPU-$\beta$ lattice~\cite{Lepri1997prl}. This is in contrast to the momentum-conserving half lattice and half HPG ding-a-ling model where normal heat conduction has been obtained~\cite{Dadswell2010pre}.
The underlying physical mechanism behind this discrepancy is an open issue and deserves further investigation in the future.

\section{acknowledgments}
The numerical calculations were carried out at Shanghai Supercomputer Center, which
has been supported by the NSF China with grant No. 11334007(B.L.). This work has
been supported by the NSF China with grant No. 11334007(Z.G., N.L., B.L.), the NSF
China with Grant No. 11205114(N.L.), the Program for New Century Excellent Talents
of the Ministry of Education of China with Grant No. NCET-12-0409(N.L.) and the
Shanghai Rising-Star Program with grant No. 13QA1403600(N.L.).

\bibliographystyle{apsrev4-1}

\end{document}